\documentstyle[prl,aps,floats,epsf,psfrag]{revtex}

\begin{document}

\twocolumn[\hsize\textwidth\columnwidth\hsize\csname
@twocolumnfalse\endcsname

\draft

\title{Nonlinear Resonant Transport of Bose Einstein Condensates}

\author{Tobias Paul, Klaus Richter, and Peter Schlagheck}

\address{Institut f{\"u}r Theoretische Physik, Universit{\"a}t Regensburg, 
         93040 Regensburg, Germany}

\date{\today}

\maketitle

\begin{abstract}

The coherent flow of a Bose-Einstein condensate through a quantum dot in a
magnetic waveguide is studied.
By the numerical integration of the time-dependent Gross-Pitaevskii equation
in presence of a source term, we simulate the propagation process of the
condensate through a double barrier potential in the waveguide.
We find that resonant transport is suppressed in interaction-induced regimes
of bistability, where multiple scattering states exist at the same chemical
potential and the same incident current.
We demonstrate, however, that a temporal control of the external potential can
be used to circumvent this limitation and to obtain enhanced transmission near
the resonance on experimentally realistic time scales.

\end{abstract}


]

\narrowtext

The rapid progress in the fabrication and manipulation of ultracold
Bose-Einstein condensates has lead to a number of fascinating experiments
probing complex condensed matter phenomena in perfectly controllable
environments, such as the creation of vortex lattices \cite{AboO01S} and the
quantum phase transition from a superfluid to a Mott insulator state in
optical lattices \cite{Gre02N}.
With the development of ``atom chips'' \cite{FolO00PRL,OttO01PRL,HaeO01N},
new perspectives are opened also towards mesoscopic physics.
The possibility to generate atomic waveguides of arbitrary complexity above
microfabricated surfaces does not only permit highly accurate matter-wave
interference experiments \cite{AndO02PRL}, but would also allow to study the
interplay between interaction and transport with an unprecedented degree of
control of the involved parameters.
The connection to electronic mesoscopic physics was appreciated by Thywissen
et al.\ \cite{ThyWesPre99PRL} who proposed a generalization of Landauer's
theory of conductance \cite{FerGoo} to the transport of non-interacting
atoms through point contacts.
Related theoretical studies were focused on the adiabatic propagation of a
Bose-Einstein condensate in presence of obstacles
\cite{LebPav01PRA,JaeSti02PRA,Pav02PRA,LebPavSin03PRA}, the dynamics of
soliton-like structures in waveguides (e.g.\ \cite{KomPap02PRL}), and the
influence of optical lattices on transport (e.g.\ \cite{HilO02PRA}), to
mention just a few examples.

Particularly interesting in this context is the propagation of a Bose-Einstein
condensate through a double barrier potential, which can be seen as a
Fabry-Perot interferometer for matter waves.
In the context of atom chips, such a bosonic quantum dot could be realized by
suitable geometries of microfabricated wires.
An alternative implementation based on optical lattices was suggested by
Carusotto and La Rocca \cite{CarLar99PRL,Car01PRA} who pointed out that the
interaction-induced nonlinearity in the mean-field dynamics would lead to a
bistability behaviour of the transmitted flux in the vicinity of resonances.
This phenomenon is well known from nonlinear optics \cite{Boy} and arises also
in electronic transport through quantum wells (e.g.
\cite{GolTsuCun87PRL,PreIonCap91PRB,Azb99PRB}) due to the Coulomb interaction
in the well.

In this Letter, we investigate to which extent resonant transport through such
a double barrier potential can be achieved for an interacting condensate 
in a realistic propagation process, where the magnetic guide is gradually
filled with matter wave.
To simulate such a process, we numerically integrate the time-dependent
Gross-Pitaevskii equation in presence of a source term that models the
coupling to a reservoir of Bose-Einstein condensed atoms.
We shall point out that resonant scattering states, which exist in principle
for arbitrarily strong interactions, cannot be populated in the above
way if the nonlinearity induces a bistability regime near the resonance.
Finally, we suggest an adiabatic control scheme that permits to circumvent
this limitation on experimentally feasible time scales.

\begin{figure}[t]
  \begin{center}
    \psfrag{potential}{$V / (\hbar \omega_\perp)$} 
    \psfrag{V}{$V$} 
    \psfrag{psi2}{$\vert\psi\vert^2$} 
    \psfrag{xcoord}{$x/a_\perp$} 
    \epsfxsize8cm
    \epsfbox{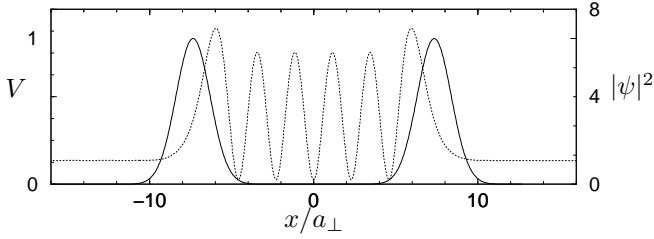}
  \end{center}
  \caption{External longitudinal potential $V$ in units of $\hbar \omega_\perp$.
    The dotted line shows the longitudinal atom density (in units of
    $a_\perp^{-1}$) of the scattering state associated with the 5th excited
    resonance, calculated at $\mu = 1.127 \hbar \omega_\perp$ and $j_t = 1.6 \omega_\perp$.
    \label{fg:potwf}
  }
\end{figure}

We consider a coherent beam of Bose-Einstein condensed atoms propagating
through a double barrier potential in a magnetic waveguide.
In presence of a strong cylindrical confinement with trapping frequency
$\omega_{\perp}$, the mean-field dynamics of the condensate is described by the
effective one-dimensional Gross-Pitaevskii equation
\begin{equation}\label{eq:gp1d}
  i\hbar\frac{\partial \psi}{\partial t} = \left(-\frac{\hbar^2}{2 m} \frac{\partial^2}{\partial x^2}
    + V(x) + g \vert \psi(x,t)\vert^{2}\right)\psi(x,t)
\end{equation}
with $g = 2 a_{s} \hbar \omega_{\perp}$ \cite{Ols98PRL}, where $m$ is the mass and $a_{s}$
the $s$-wave scattering length of the atoms.
For the sake of definitness, the double barrier potential is given by
\begin{equation}
  V(x) = V_{0}\left[ e^{-(x+L/2)^2/\sigma^2 }+ e^{-(x-L/2)^2/\sigma^2 }\right]
\end{equation}
(see Fig.~\ref{fg:potwf}).
Our numerical calculations were performed for $^{87}$Rb atoms ($a_{s}=5.77$nm)
with $\omega_{\perp}= 2 \pi \times 10^{3} $s$^{-1}$, $a_\perp = \sqrt{\hbar/m\omega_\perp} \simeq 0.34\mu$m, 
$V_{0}= \hbar\omega_{\perp}$, and $L = 10 \sigma = 5 \mu$m $ \simeq 14.7 a_\perp$ \cite{remark}. 
This yields $g \simeq 0.034 \hbar \omega_{\perp} a_\perp$.

Let us first discuss resonances in terms of stationary scattering states of
the condensate.
The latter are given by stationary solutions $\psi(x,t) = \psi(x) \exp(- i \mu t / \hbar
)$ of Eq.~(\ref{eq:gp1d}) satisfying outgoing boundary conditions of the form
$\psi(x) = A e^{i k x}$ with $k > 0$ for $x \to \infty$.
To calculate them, we insert the ansatz $\psi(x) = A(x)\exp\left[i \phi(x) \right]$
(with real $A$ and $\phi$) into the stationary Gross-Pitaevskii equation, and
separate the latter into real and imaginary parts. 
This yields the condition that the current $j(x)=(\hbar / m) A^2(x) \phi'(x) \equiv j_t$
is independent of $x$, and
\begin{equation}\label{eq:ampl}
  \mu A = -\frac{\hbar^2}{2 m}\frac{d^2A}{d x^2}+\left(V(x)+ \frac{m}{2}
  \frac{j_{t}^2}{A^4}\right)A + g A^3
\end{equation}
as equation for the amplitude $A(x)$.
The latter can be numerically integrated from the ``downstream'' ($x \to \infty$) to
the ``upstream'' ($x \to -\infty$) region by means of a Runge-Kutta solver.
As ``asymptotic condition'' at $x \to \infty$, we choose $A' = 0$ and $A$
satisfying
\begin{equation}\label{eq:mu}
  \mu = \frac{m}{2} \frac{j_{t}^2}{n^2} + g n,
\end{equation}
for a given $j_t$, with $n \equiv A^2$ the longitudinal density of the condensate.
As was pointed out in Ref.~\cite{LebPav01PRA}, Eq.~(\ref{eq:mu}) exhibits a
low-density (supersonic) and a high-density (subsonic) solution, where the
transport is respectively dominated by the kinetic energy and by the mutual
interaction of the atoms.
Since in realistic propagation processes the waveguide is initially empty in
the downstream region, we choose the low-density solution for the asymptotic
value of $A$.

\begin{figure}[t]
  \begin{center}
    \psfrag{jt}{$j_t$}
    \psfrag{mu/hbarom}{$\mu / (\hbar \omega_\perp)$}
    \epsfxsize8cm
    \epsfbox{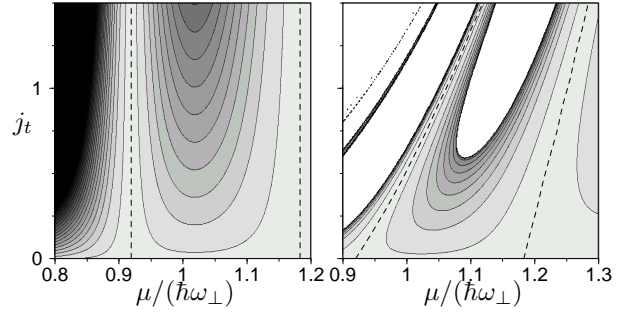}
  \end{center}
  \caption{Drag exerted by the condensate onto the obstacle, plotted as a
    function of the chemical potential $\mu$ and the total current $j_t$ (in
    units of $\omega_\perp$) for $g=0$ (left panel) and $g=0.034 \hbar \omega_{\perp} a_\perp$ (right
    panel).
    Light gray areas correspond to a low and dark gray areas to a high drag.
    In white areas, the integration of Eq.~(\ref{eq:ampl}) leads to a
    diverging $A(x)$.
    The location of the 5th and 6th excited resonance, where the drag
    vanishes, is marked by dashed lines.
    \label{fg:drag}
  }
\end{figure}

A measure of the proximity of the scattering state to a resonant state is
provided by the drag
\begin{equation}\label{eq:drag}
  F_{d}=\int_{-\infty}^{+\infty}dx ~n(x)~\frac{d V(x)}{dx}
\end{equation}
that the condensate exerts onto the obstacle \cite{Pav02PRA}.
Far from any resonance, the amount of reflection from the potential is rather
large, leading to a finite drag due to the associated momentum transfer,
while a vanishing drag is expected near a resonance where the condensate is
perfectly transmitted through the quantum dot.
In Fig.~\ref{fg:drag}, the drag is plotted as a function of $\mu$ and $j_t$ in
the vicinity of the 5th and 6th excited resonance, which have five and six
nodes within the well, respectively (see Fig.~\ref{fg:potwf}).
While the chemical potential of the resonant state is independent of $j_t$ in
the linear case, it increases with $j_t$ in presence of a repulsive atom-atom
interaction.
Note that for large currents scattering states with non-diverging amplitude
$A(x)$ exist only in the immediated vicinity of resonances.

Can resonant scattering states be populated in a realistic experiment where
the condensate is initially confined in a microtrap and then released to
propagate through the waveguide?
To address this question, we numerically integrate the time-dependent
Gross-Pitaevskii equation
\begin{eqnarray}
  i\hbar\frac{\partial}{\partial t} \psi(x,t) & = & \left( - \frac{\hbar^2}{2m} \frac{\partial^2}{\partial x^2} + V(x) 
    -g\vert \psi(x,t) \vert^2\right)\psi(x,t) \nonumber \\
  & & + S_0\exp(-i\mu t / \hbar) \delta(x-x_0) \label{eq:gps} 
\end{eqnarray}
in presence of an inhomogeneous source term emitting coherent matter waves
with chemical potential $\mu$ at position $x_0$ (we used $x_0 = -15 a_\perp$ in our
calculation).
The wavefunction $\psi(x,t)$ is expanded on a lattice (within $ -20 \leq x/a_\perp \leq 20$)
and propagated in real time domain.
To avoid artificial backscattering from the boundaries of the lattice, we
impose absorbing boundary conditions which are particularly suited for
transport problems \cite{Shi91PRB} and can be generalized to account for weak
or moderate nonlinearities \cite{PauSchRic_prep,cap}.

Stationary scattering states can now be calculated by propagating $\psi(x,t)$ in
presence of an adiabatic increase of the source amplitude $S_0$ from $0$ up to
a given maximum value, with the initial condition $\psi \equiv 0$.
This approach simulates a realistic propagation process where a coherent beam
of Bose-Einstein condensed atoms with chemical potential $\mu$ is injected into
the guide from a reservoir.
Furthermore, it provides a straightforward access to the transmission
coefficient $T$ that is associated with a given scattering state:
$T$ can be defined by the ratio of the current $j_t$ in {\em presence} of the
double barrier potential (i.e., the transmitted current) to the current $j_i$
obtained in {\em absence} of the potential (the incident current).
The latter is analytically evaluated as $j_i = \hbar |S_0|^2 / (m k_0)$ with $k_0$
being self-consistently defined by $k_0 = \sqrt{2 m (\mu - g |S_0|^2 / k_0^2)} /
\hbar$.

\begin{figure}[t]
  \begin{center}
    \psfrag{T}{$T$}
    \psfrag{mu/hbarom}{$\mu / (\hbar \omega_\perp)$}
    \epsfxsize8cm
    \epsfbox{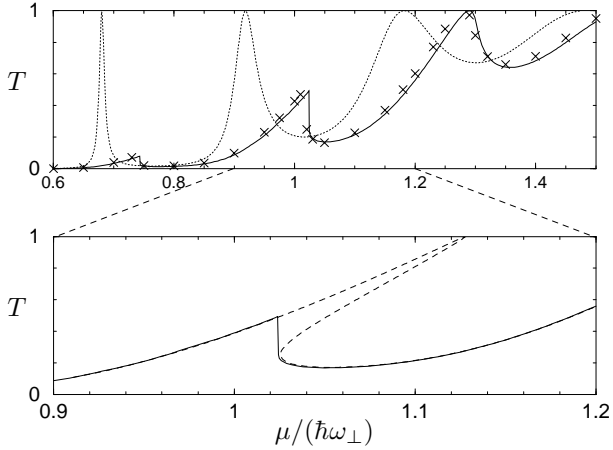}
  \end{center}
  \caption{Transmission spectrum obtained from the time-dependent propagation
    approach for $g = 0.034 \hbar \omega_{\perp} a_\perp$ and fixed incident current $j_i = 1.6
    \omega_\perp$ (solid line) compared, in the upper panel, to the interaction-free
    spectrum (dotted line).
    The crosses show the results of a full three-dimensional calculation
    \protect\cite{PauSchRic_prep}.
    The dashed line in the lower panel shows the two other branches of the
    5th resonance peak, which are not populated by the time-dependent
    propagation process.
    \label{fg:spectrum}
  }
\end{figure}

Fig.~\ref{fg:spectrum} shows the transmission coefficient as a function of the
chemical potential around the 5th excited resonance.
For each value of $\mu$, the wavefunction was propagated according to
Eq.~(\ref{eq:gps}) in presence of an adiabatic increase of the source
amplitude $S_0$ up to the maximum value that corresponds to the incident
current $j_{i} = 1.6 \omega_\perp$.
The transmitted current is directly evaluated from the stationary scattering
state obtained at the end of the propagation procedure.
While the typical sequence of Breit-Wigner resonances is obtained in the
linear case (or in the limit of very low incident currents), the profiles
become asymmetric for $g > 0$ with perfect transmission being suppressed for
narrow resonances.
These results are essentially reproduced by a full three-dimensional
mean-field calculation, which will be described elsewhere
\cite{PauSchRic_prep}.

The step-like structures in the transmission spectrum indicate a 
bistability phenomenon, as known from similar processes in nonlinear
optics \cite{Boy} and in electronic transport through quantum wells
(e.g.\ \cite{GolTsuCun87PRL,PreIonCap91PRB,Azb99PRB}).
Additional branches of the resonance peaks are indeed identified by the
integration method based on Eq.~(\ref{eq:ampl}) which allows to calculate
stationary scattering states for given $j_t$ and $\mu$.
The incident current of the scattering state is approximately determined
according to Ref.~\cite{LebPavSin03PRA} via
\begin{equation}\label{eq:j}
  j_i \simeq \sqrt{\frac{\mu - g n_{\rm av}}{2 m}} 
  \left( n_{\rm av} + \sqrt{n_{\rm max} n_{\rm min}} \right)
\end{equation}
with $n_{\rm av} = \frac{1}{2}(n_{\rm max} + n_{\rm min})$, where 
$n_{\rm max}$ and $n_{\rm min}$ denote the maxima and minima, respectively, of
the longitudinal upstream density.
The expression (\ref{eq:j}) assumes a cosine-like oscillation of the upstream
density, and is valid for small $g (n_{\rm max} - n_{\rm min}) / n_{\rm av}$.

Finding the value of $j_t$ that results from a given $j_i$ is now an
optimization problem that can be solved straightforwardly.
The result is shown in the lower panel of Fig.~\ref{fg:spectrum} for chemical
potentials around the 5th resonant state.
In addition to the spectrum obtained by the integration Eq.~(\ref{eq:gps}),
two further solutions appear for $1.02 < \mu/(\hbar\omega_\perp) < 1.13$ which join together
to form a resonance peak that is asymmetrically distorted towards higher $\mu$.
The existence of such a multivalued spectrum, which is reminiscent of
nonlinear oscillators, was in this context pointed out by Carusotto and La
Rocca \cite{CarLar99PRL}.
Since the additional branches of the resonance peak are apparently not
populated by the time-dependent integration approach, we expect that resonant
transport will generally be suppressed in a realistic propagation process.
Loosely speaking, the atoms ``block'' each other when going through the double
barrier potential \cite{block}.

To enhance the transmission of matter waves near a narrow resonance, the
external potential needs to be varied {\em during the propagation process}.
Specifically, this can be achieved e.g.\ by illuminating the scattering region
with a red-detuned laser pulse.
We can describe such a process by a temporal modulation of $V$
according to 
\begin{equation}
  V(x) \longrightarrow V(x,t) \equiv V(x) - V_0(t) \label{eq:pot1}
\end{equation}
where $V_0(t) > 0$ is determined by the detuning and the time-dependent
intensity of the laser.
For an adiabatic modulation of $V$, the wavefunction $\psi(x,t)$ will, at each
time $t$, remain close to the instantaneous scattering state associated with
the external potential (\ref{eq:pot1}) --- or, equivalently formulated, to the
scattering state for $V = V(x)$ at the shifted chemical potential 
$\mu + V_0(t)$.
Outside the bistability regime, e.g.\ for $\mu + V_0(t) < 1.02 \, \hbar\omega_\perp$ in case of
the 5th resonance, this scattering state would be uniquely given by the one
that is also obtained by the direct propagation process.
However, as soon as the effective chemical potential $\mu + V_0(t)$ is raised
above $1.02 \, \hbar\omega_\perp$, the wavefunction will continuously
{\em follow the upper branch} of the resonance and evolve into a 
near-resonant scattering state with high transmission.

\begin{figure}[t]
  \begin{center}
    \psfrag{T}{$T$}
    \psfrag{V/hbarom}{$V_0/(\hbar\omega_\perp)$}
    \psfrag{t/om1}{$\omega_\perp t$}
    \epsfxsize8cm
    \epsfbox{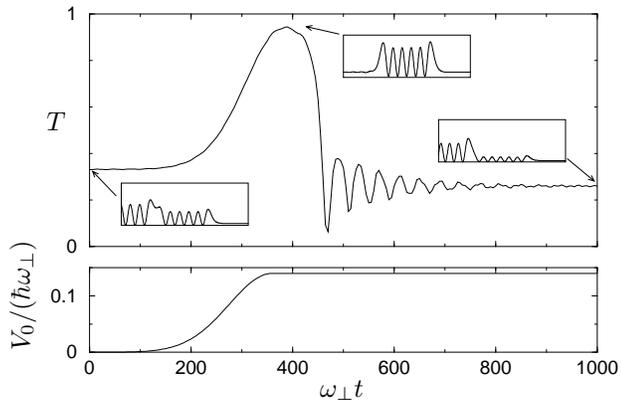}
  \end{center}
  \caption{Time evolution of the transmission coefficient $T$ (upper panel)
    during the ramping process of the external potential $V_0$ (lower panel)
    which shifts the effective chemical potential from $\mu = 0.985$ to 
    $1.125 \hbar \omega_\perp$.
    As shown in the insets (with scaling as in Fig.~\ref{fg:potwf}), the
    wavefunction adiabatically evolves into a nearly resonant state with
    transmission close to unity, and decays from there to the low-transmission
    scattering state within a time scale of the order of $\tau \sim 100 \omega_\perp^{-1}$. 
    \label{fg:control}
  }
\end{figure}

Indeed, we can use our numerical setup to simulate such a process.
Fig.~\ref{fg:control} shows the transmission coefficient as a function of the
propagation time, where the effective chemical potential was shifted from 
$\mu = 0.985 \, \hbar\omega_\perp$ to $\mu + V_0^{\rm max} = 1.125 \, \hbar\omega_\perp$ by means of a
Gaussian ramping process taking place within $0 < \omega_\perp t < 360$ (see the lower
panel of Fig.~\ref{fg:control}).
We see that the transmission approaches unity at the end of the ramping
process, which clearly indicates that the scattering wavefunction evolves
along the upper branch of the resonance peak.
This is indeed confirmed by the associated density shown in the insets
(to be compared with Fig.~\ref{fg:potwf}).

As was also pointed out in the context of electronic transport through quantum
wells \cite{Azb99PRB}, the resonant scattering state is dynamically unstable
in presence of interactions.
This instability is indeed encountered in our system:
Continuing the numerical propagation beyond $\omega_\perp t = 360$ (at fixed $V_0$)
results in a decay of the wavefunction towards a low-transmission scattering
state within a time scale of the order of $\tau \sim 100 \omega_\perp^{-1} \simeq 16$ms.
This lifetime should be long enough, however, to transport a large fraction of
the condensate through the double barrier, as well as to manipulate the
resonant scattering state: 
by closing, for instance, the potential well during that time scale (e.g.\
with blue-detuned lasers that enhance the potential outside the barriers), one
would create a trap in which an interacting mean-field state with an unusually
high excitation (with five nodes in case of the 5th excited resonance) would
be obtained.

In conclusion, we have studied resonant transport of interacting Bose-Einstein
condensates through a symmetric double barrier potential in a magnetic
waveguide.
The nonlinearity induced by the interaction leads to a distortion of the
resonance peak, where the associated scattering state cannot be populated by
directly sending coherent matter wave onto the initially empty waveguide.
To obtain nevertheless a finite amount of transmission on intermediate time
scales, the external potential needs to be adiabatically varied during the
propagation process.
The lifetime of the resonant scattering state obtained in this way is
predicted to be of the order of $\tau \sim 10$ms for our particular setup, which
should be long enough to allow for further experimental manipulations.
We expect that the basic principles of the scenario encountered for our double
barrier potential apply also to more complex quantum dot geometries such as
sequences of more than two barriers along the guide.
This indicates that the design of suitable control schemes will be a relevant
issue for the mesoscopic transport of Bose-Einstein condensates.

It is a pleasure to thank Nicolas Pavloff, Peter Schmelcher, Joachim Brand,
J{\'o}zsef Fort{\'a}gh, and Wilhelm Prettl for fruitful and inspiring discussions.


\begin{thebibliography}{10}
\bibitem{remark}
Such a small barrier width would be at the limit of realizability with
  present-day atom-chip technology. We remark, however, that the phenomena we
  discuss here do not sensitively depend on the chosen parameters nor on the
  specific shape of the potential.
\bibitem{cap}
We reproduced our results by an alternative approach using complex absorbing
  potentials at the boundaries.
\bibitem{block}
This phenomenon must not be confused with the "atom blockade" discussed by
  Carusotto \cite{Car01PRA}, which arises in the limit of few atoms and very
  strong interactions.
\end{thebibliography}
\end{document}